\documentclass[aps,prb,twocolumn,showpacs,floatfix]{revtex4}
\usepackage{amssymb}
\usepackage{dcolumn}
\usepackage{amstext}
\usepackage{graphicx}
\usepackage{amsmath}
\usepackage[normalem]{ulem}
\usepackage{color}

\begin{document}

\title{A strong coupling critique of spin fluctuation driven charge order in underdoped cuprates}
\author{Vivek~Mishra}
\altaffiliation{Present address : Center for Nanophase Materials Sciences, Oak Ridge National Laboratory, Oak Ridge, TN 37831}
\affiliation{Materials Science Division, Argonne National Laboratory, Argonne, IL 60439}
\author{M.~R.~Norman}
\affiliation{Materials Science Division, Argonne National Laboratory, Argonne, IL 60439}

\date{\today}

\begin{abstract}
Charge order has emerged as a generic feature of doped cuprates, leading to important
questions about its origin and its relation to superconductivity.
Recent experiments on two classes of hole doped cuprates
indicate a novel d-wave symmetry for the order.  These were motivated by
earlier spin fluctuation theoretical studies based on an expansion about hot spots
in the Brillouin zone that indicated such order would be competitive with d-wave superconductivity.
Here, we reexamine this problem by solving strong coupling equations in the full
Brillouin zone for experimentally relevant parameters.  We find that bond-oriented order, as seen experimentally,
is strongly suppressed. We also include coupling to B$_{1g}$ phonons and do not see any qualitative change.
Our results argue against an itinerant model for the charge order, implying instead that such order
is likely due to Coulombic phase separation of the doped holes.
\end{abstract}
\pacs{74.72.-h, 75.25.Dk, 71.45.Lr, 74.20.Mn}
\maketitle
Following earlier NMR studies \cite{Wu2011},
recent x-ray experiments on underdoped cuprates 
have detected short range charge density wave (CDW) order with a period of 3 to 5 lattice spacings in
YBCO \cite{Ghiringhelli2012,Chang2012,Achkar2012}, and in Bi based \cite{Comin2014,Neto2014}  
and Hg based cuprates \cite{Tabis2014}.
For YBCO, the doping range where the charge order has been observed \cite{Canosa2014}
coincides with the doping range where quantum oscillation experiments detect reconstruction
of the Fermi surface \cite{SebastianReview}.
In conventional CDW systems,
the charge order is thought to have s-wave symmetry \cite{GrunerRMP}.
In contrast, 
scanning tunneling microscopy \cite{Fujita2014} and  
resonant soft x-ray scattering \cite{Comin2015,Achkar2015} data have
revealed a novel d-wave symmetry, where the two oxygen ions in a CuO$_2$ unit are out of phase.
This charge order differs from the more robust 
stripe order seen earlier in La based compounds \cite{Tranquada1995,Kim2008,Hucker2011},
which appears to have s-wave symmetry instead \cite{Achkar2015}.
In all cases, though, the wavevector is oriented along the bond direction \cite{foot1}.

\begin{figure}[b]
\includegraphics[width=.95\columnwidth]{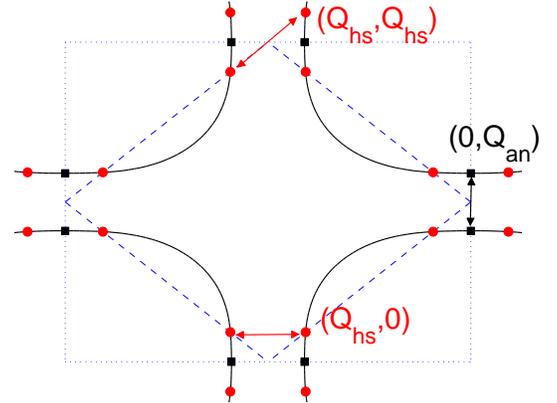}
\caption{(Color online) Fermi surface for the tight binding
dispersion considered in this work. The dashed lines show the
antiferromagnetic Brillouin zone and the dotted lines the
structural one.  Filled circles denote
the hot spots and filled squares the antinodal points.
The wavevectors studied here are indicated by the arrows.}
\label{FS_cartoon}
\end{figure}

The search for d-wave symmetry was
motivated by earlier theoretical studies of Metlitski and Sachdev \cite{Metlitski2010,Metlitski2010b}.
They have shown that charge order is competitive with d-wave superconductivity
in a spin fluctuation model. This instability has a d-wave form factor, with
a diagonal wavevector that spans Fermi surface points (hot spots) that intersect the antiferromagnetic
zone boundary of the undoped phase (Fig.~\ref{FS_cartoon}).
The subsequent observation of charge order in YBCO motivated a number of follow up 
studies \cite{Pepin2013,LaPlaca2013,Sau2014,Allais2014,WangChubukov}.
Most of these studies are either based on an expansion around the hot spots, 
with the Fermi surface curvature treated
as a perturbation \cite{Pepin2013,Sau2014,WangChubukov}, or rely on a weak coupling approximation.
So, the question arises whether these results survive in a strong coupling treatment where these approximations are not made. 

Here, we solve the strong coupling instability equation for the charge order in the entire Brillouin zone
including the full momentum and frequency dependence of the bosonic and fermionic spectra.
This formalism
has been used in the past to study d-wave superconductivity originating from spin 
fluctuations \cite{Pines1992,ScalapinoRMP}.
It has also been used to study instabilities in the particle-hole channel \cite{Bulut1993}. 
Recently, this formalism was used by us to examine the effect of the pseudogap on
spin-fluctuation mediated pairing \cite{MCCN2014}.
The one approximation we make is that the bosonic and fermionic spectra are taken from experiment rather than self-consistently calculated.
We find that bond-centered charge order is completely suppressed, and inclusion of the dressed fermion 
Green's function additionally suppresses diagonal charge order, with the only robust order in this model
for experimentally relevant parameters being d-wave superconductivity.

Our starting point is the linearized equation for 
the anomalous self energy in the particle-hole channel
\begin{eqnarray}
 &&T \sum_{k^\prime,\omega_m} V(k-k^\prime,i\omega_n -i \omega_m) \mathcal{P}^{Q}(k^\prime,i\omega_m) \Phi^{Q}(k^\prime, i \omega_m) \nonumber \\
&=& \lambda \Phi^{Q}(k,i\omega_n).
 \label{eq:cdw}
\end{eqnarray}
Here $Q$ is the ordering vector, $V$ is the interaction and $\mathcal{P}^{Q}$
is the CDW particle-hole kernel
\begin{equation}
 \mathcal{P}^{Q}(k^\prime,i\omega_m)=G(k^\prime-\frac{Q}{2} , i\omega_m )G(k^\prime+\frac{Q}{2} , i\omega_m )
\end{equation}
where $G$ is the fermion Green's function.
Because of the complexity of the strong coupling equations,
as a first approximation, we do not include the frequency dependence of the fermion self-energy,
and thus set $G(k,i\omega)=(i\omega-\xi_k)^{-1}$.
For $\xi_k$, we consider a renormalied dispersion that fits low energy ARPES data
for Bi$_2$Sr$_2$CaCu$_2$O$_{8+\delta}$ 
(tb2 dispersion of Ref.~\onlinecite{MRN2007}).
Later in the paper, we will include the fermion self-energy as well.
The interaction assumed here is
 \begin{equation}
 V(k,i\Omega_n)=  \int_{-\infty}^{\infty}\frac{dx}{\pi}\frac{V^{\prime\prime}(k,x)}{i\Omega_n-x}
\end{equation}
where $V^{\prime\prime}$ is proportional to the imaginary part of the dynamic spin susceptibility.
We consider the phenomenological form \cite{MMP1990}
\begin{eqnarray}
V^{\prime\prime}(k,\Omega) &=& \frac{3}{2} g_{sf}^2 \chi_Q \frac{ \Omega \Omega_{sf} }{\chi_k^2 \Omega_{sf}^2 + \Omega^2 } \nonumber \\
\chi_k &=& (\xi_{AF}/a)^{-2}+2 + \cos k_xa + \cos k_ya
\end{eqnarray}
where $g_{sf}$ is the spin-fermion coupling constant, $\xi_{AF}$ is the antiferromagnetic correlation length, $\chi_Q$ is
the static susceptibility at $Q_{AF}=(\pi/a,\pi/a)$ and $\Omega_{sf}$ is the characteristic spin-fluctuation energy scale.
Because of the 1/$\Omega$ decay of $V^{\prime\prime}$, we
impose a frequency cut-off, $\Omega_c$.
We use $\Omega_c=$300 meV, $\Omega_{sf}=$100 meV, $g^2_{sf}\chi_Q=$0.9 eV and $\xi_{AF}=2a$, where $a$ is the lattice constant.
The values of $\xi_{AF}$ and $\Omega_{sf}$ are motivated from inelastic neutron scattering studies of the magnetic excitations of underdoped YBCO
near $Q_{AF}$ \cite{Fong00,Dai01}.
$\Omega_c$ is motivated from recent resonant inelastic x-ray scattering studies of the higher energy excitations away 
from $Q_{AF}$ \cite{Tacon2011,Dean2013}.  The value of $g^2_{sf}\chi_Q$ was chosen to obtain a superconducting T$_c$ of 50 K, typical
for underdoped YBCO.
 At the transition, the leading eigenvalue $\lambda$ in Eq.~(\ref{eq:cdw}) reaches 1, with its eigenvector
giving the structure of the CDW order parameter.

\begin{figure}[h]
\includegraphics[width=.95\columnwidth]{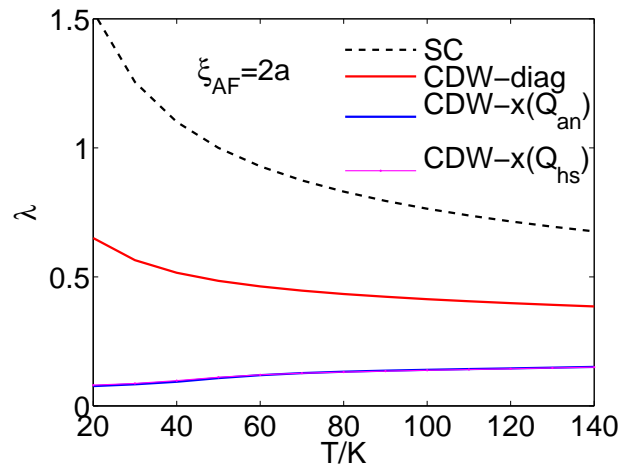}
\caption{(Color online) Temperature dependence of the leading eigenvalues
for the superconducting d-wave state (SC), the diagonal 
CDW state (CDW-diag) and the bond oriented CDW state (CDW-x),
in an approximation where $G$ is based on a renormalized dispersion taken from ARPES data.
The ordering vector for CDW-diag is ($Q_{hs},Q_{hs}$).
For CDW-x, two vectors were considered:
($Q_{an},0$) and ($Q_{hs},0$).}
\label{Tdep:eigenvalue}
\end{figure}
We consider the CDW instabilities for the diagonal CDW case (CDW-diag) with ordering vector ($Q_{hs},Q_{hs}$),
and for the bond oriented case (CDW-x) with vectors ($Q_{an},0$) and ($Q_{hs},0$),
as shown in Fig.~1.
We perform our calculations with a 0.02$\pi/a$ momentum grid with 32 Matsubara frequencies, which is sufficient 
for convergence of the eigenvalues for the temperature range studied here (see supplementary information).
Fig.~\ref{Tdep:eigenvalue} shows the temperature dependence of the leading
eigenvalue for the different CDW cases along with the d-wave superconducting case.

As expected, the eigenvalue for the superconducting case exhibits a logarithmic divergence in $T$.
This is present as well for CDW-diag order, though we find it to be significantly reduced relative
to the superconducting one.
\begin{figure*}
\includegraphics[width=0.95\columnwidth]{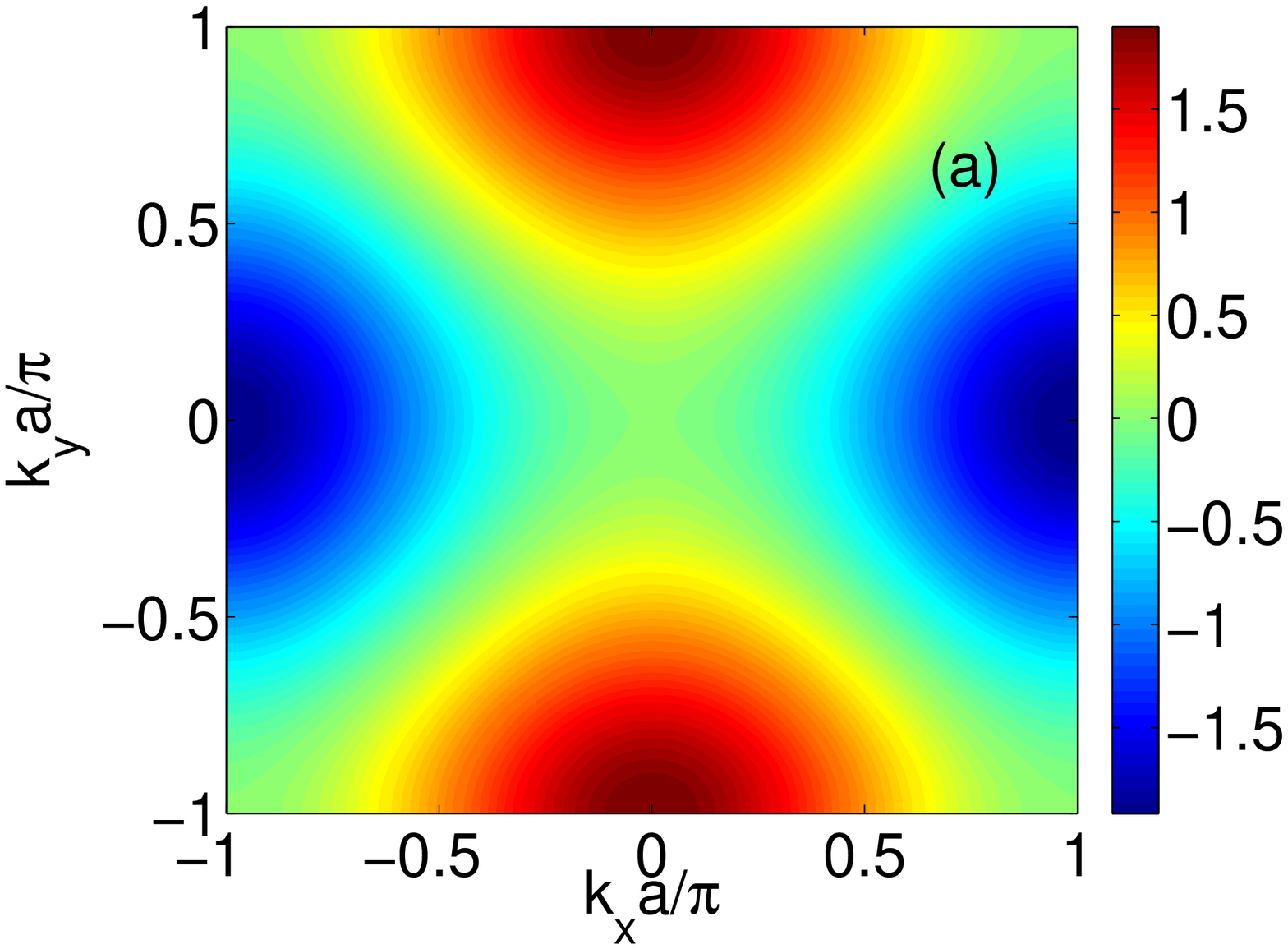}
\includegraphics[width=0.92\columnwidth]{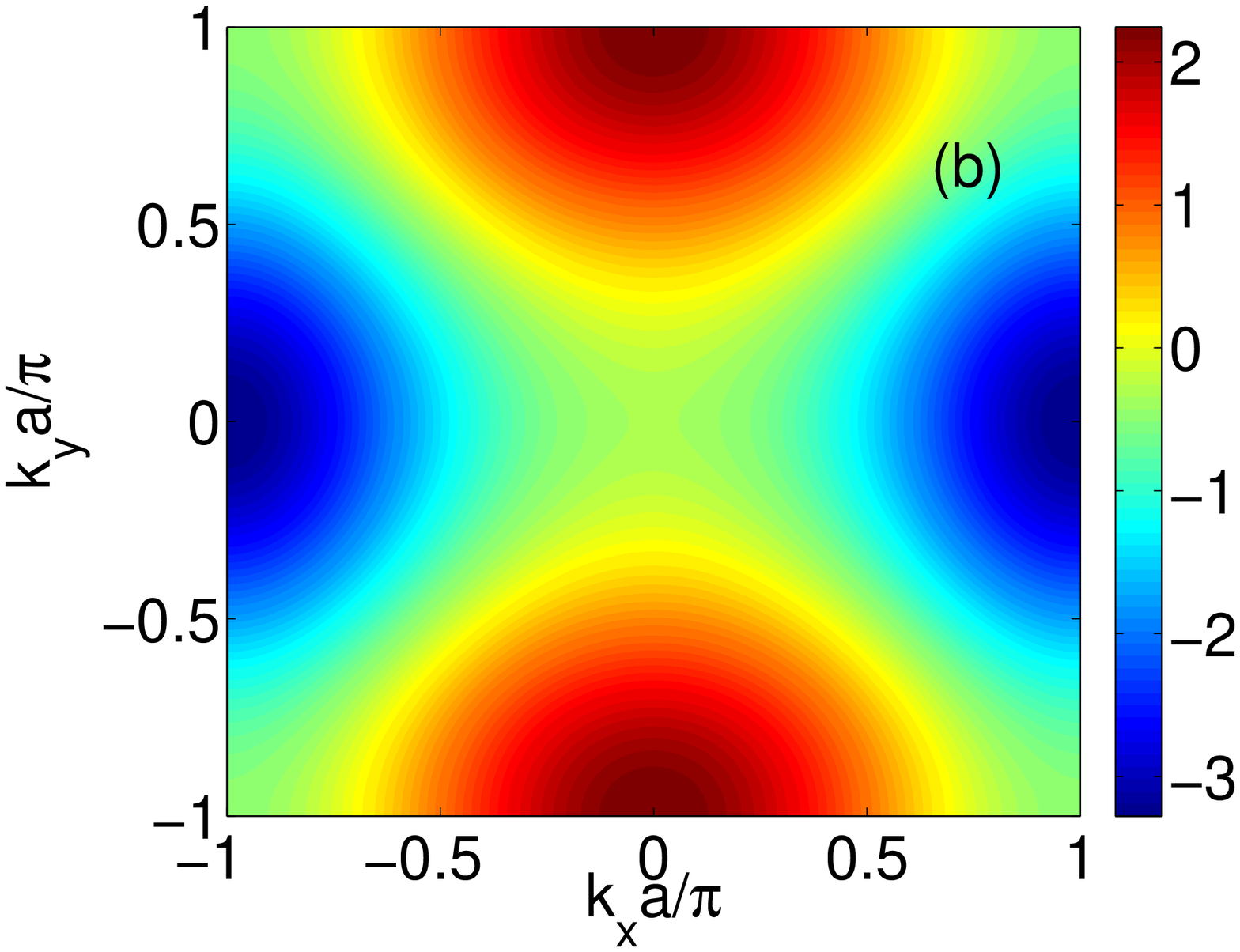}
\caption{(Color online) The momentum dependence of the eigenvectors at the lowest
Matsubara frequency ($T$=40 K) corresponding to leading
eigenvalues (Fig.~2) for the (a) CDW-diag and (b) CDW-x states. 
For CDW-x, the ordering vector is ($Q_{an},0$), though the eigenvector for ($Q_{hs},0$) is similar.
A normalization condition of $T\sum_{\omega_n,k} |\Phi^Q(k,i\omega_n)|^2 = 1$ is employed.}
\label{Evec:bondxy}
\end{figure*} 
The CDW-diag state has d-wave (B$_{1g}$) symmetry with a momentum
dependence that is well described by $\cos(k_xa)-\cos(k_ya)$ as can be
seen in Fig.~\ref{Evec:bondxy} (a).
Increasing the antiferromagnetic coherence length doesn't change our findings.
The CDW-diag instability becomes stronger with longer $\xi_{AF}$, but
it always remains subdominant relative to d-wave superconductivity (see supplementary material).

We now focus on the bond-oriented CDW states, since there is no experimental evidence for diagonal-oriented order.
The $T$ dependence of the leading eigenvalues for CDW-x are also plotted in Fig.~\ref{Tdep:eigenvalue}.
They are almost identical for vectors ($Q_{hs},0$) and ($Q_{an},0$).
The eigenvalues vary little with temperature, showing no evidence for a log in the temperature range
studied.  This is one of our main results, and
differs from an analytic approximation quoted in earlier work \cite{WangChubukov}.
The lack of a log divergence in our case is likely due to the 
curvature coming from the experimentally determined Fermi surface.  Moreover, our bosonic
spectrum is based on inelastic neutron and x-ray scattering measurements. Hence, we are not in the extreme 
quantum critical regime considered in Ref.~\onlinecite{WangChubukov}.

We next look at the structure of the CDW-x state.
Fig.~\ref{Evec:bondxy}
(b) shows the momentum dependence of the eigenvector corresponding to the leading eigenvalue,
and can be well fit by a sum of a constant (s), $\cos(k_xa)+\cos(k_ya)$ (s$^\prime$) and $\cos(k_xa)-\cos(k_ya)$ (d),
with the d-wave component dominant, consistent with earlier studies \cite{LaPlaca2013}.
\begin{figure}[h]
\includegraphics[width=.95\columnwidth]{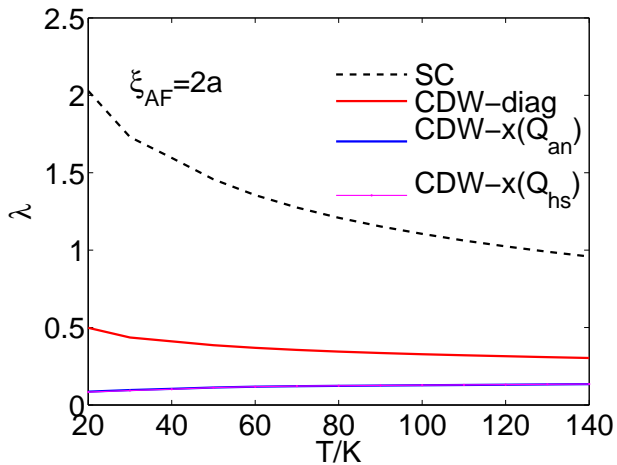}
\caption{(Color online) Temperature dependence of the leading eigenvalues
for the superconducting d-wave state (SC), the diagonal 
CDW state (CDW-diag) and the bond oriented CDW state (CDW-x), obtained using dressed Green's
functions.}
\label{Tdep:eigenvalueFullG}
\end{figure}

We now turn to the influence of the fermion self-energy. The fermion self-energy in a lowest order approximation is
\begin{eqnarray}
 \Sigma(k,i\omega) &=& T \sum_{k^\prime,\omega^\prime} V(k-k^\prime,i\omega-i\omega^\prime) G_0(k^\prime,i\omega^\prime)  \label{eq:SE1} \\
 &=& T \sum_{k^\prime,\omega^\prime} V(k-k^\prime,i\omega-i\omega^\prime) \frac{-i\omega^\prime-\xi_{k^\prime}}{\omega^{\prime2}+\xi^2_{k^\prime}}
 \label{eq:SE2}
\end{eqnarray}
where $G_0$ is the bare Green's function. Since our previous $\xi_k$ was based on ARPES data, we scale it by a factor of two to get an approximation to the bare dispersion, with a bare nodal Fermi velocity $v_{F}^{node}=$3.2eV-{\AA}.
For the spin-fluctuation interaction, we use $g^2_{sf}\chi_Q=3.2$eV and keep the
rest of the parameters the same as before. This coupling strength renormalizes the nodal Fermi velocity 
to $1.8$eV-{\AA}, which is comparable to the experimental value. To keep the shape of the Fermi surface intact,
we drop the term proportional to $\xi_{k^\prime}$ in the self-energy calculation (Eq.~\ref{eq:SE2}). We now use
this dressed Green's function ($G^{-1}(k,i\omega)=i\omega - \xi_k -\Sigma(k,i\omega)$) for
the instability analysis in Eq.~\ref{eq:cdw}.
Its effect (Fig.~\ref{Tdep:eigenvalueFullG}) is to additionally suppress the diagonal charge order
(presumably due to the energy smearing of $G$).
As a consequence, for experimentally relevant parameters,
only d-wave superconductivity remains as a robust instability in this model.

Our results cast doubt on an itinerant spin fluctuation mediated origin for the observed charge
order.  On the other hand, the dependence of the observed wavevector on
doping \cite{Canosa2014} is suggestive that the Fermi surface is playing
some role as in classic CDW systems.  This is in contrast to the La based cuprates whose
doping dependence is opposite to this, as would be expected from a real space picture
where the wavevector is proportional to the doping.  In classic CDW systems like {\it 2H}-NbSe$_2$,
phonons play an important role \cite{Weber2011}. Hence, one could think that
this might be the case in the cuprates as well, where anomalies have been seen
in both optic \cite{ReznikReview} and acoustic \cite{LeTacon2013} phonon modes near the
charge ordering vector.
It is interesting to note that $B_{1g}$ phonon modes have been postulated to be
responsible for dispersion anomalies seen in photoemission near the antinodes \cite{Cuk2004},
and perhaps their d-wave symmetry is related to that of the charge order.
In support of this, several theoretical
studies have suggested that coupling of the electrons to such modes can cause d-wave charge order \cite{Fu06,Newns07}.
We examine this idea by including the phonon mediated interaction along with the spin-fluctuation mediated interaction
in the CDW instability equation.  Under the assumption that the electron-phonon interaction is modest,
we do not find that its addition has sufficient strength
to cause a CDW instability either (see supplementary material).

None of the above means that spin mediated interactions are irrelevant for the charge order,
but our results indicate that an itinerant treatment of the problem in a parameter range appropriate to
experiment is not adequate to capture the resulting
physics.  In that context, we point to real space treatments of the $t-J$ model like
density matrix renormalization group (DMRG) 
which indicate a strong tendency to bond-centered charge order in the underdoped regime \cite{white},
with the d-wave structure likely due to Coulomb repulsion which acts to enforce charge neutrality in each unit cell.

{\it Note added in proof.}  After the original submission of this work, a critique of it has been offered by
Wang and Chubukov \cite{wang2,wang3}.  Their primary criticism revolves around two points - that
in the original submission, we neglected the fermion self-energy that they argue lessens the detrimental impact of the Fermi
surface curvature, and that our value of $\Omega_{sf}$ is too small by an order of magnitude.
In regards to the first point,
using dressed Green's functions, 
we now find strong suppression of the charge order, even for diagonal wavevectors.
In regards to the second point, our value of $\Omega_{sf}$ is based on experiment, and an order
of magnitude larger value would lead to a spin fluctuation energy scale at the $\Gamma$ point
of the zone of 4 eV, in gross disagreement with inelastic x-ray scattering data.
We also comment that the logarithmic temperature divergence of the eigenvalue derived for bond
centered charge order in their work is based on analytic approximations which we feel are not valid for
experimentally realistic parameters.  Certainly, we find no evidence for a log in our own numerical
studies, even for a $\xi_{AF}$ as large as 10$a$ (see supplementary material). 
As an additional note, our results are consistent with 
recent findings based on a real-space version of Eliashberg theory \cite{BauerSachdev}.

This work was supported by the Center for Emergent
Superconductivity, an Energy Frontier Research Center funded by the
US DOE, Office of Science, under Award No.~DE-AC0298CH1088.
We gratefully acknowledge the computing resources provided on Blues and
Fusion, the high-performance computing clusters operated by the Laboratory
Computing Resource Center at Argonne National Laboratory.
\appendix

\clearpage
\linespread{1}
\setcounter{figure}{0}
\renewcommand{\figurename}{Supplementary Fig.}
\section*{Supplementary material}
\linespread{1}

\section*{Effect of a larger antiferromagnetic coherence length}
\begin{figure}[h]
\includegraphics[width=.95\columnwidth]{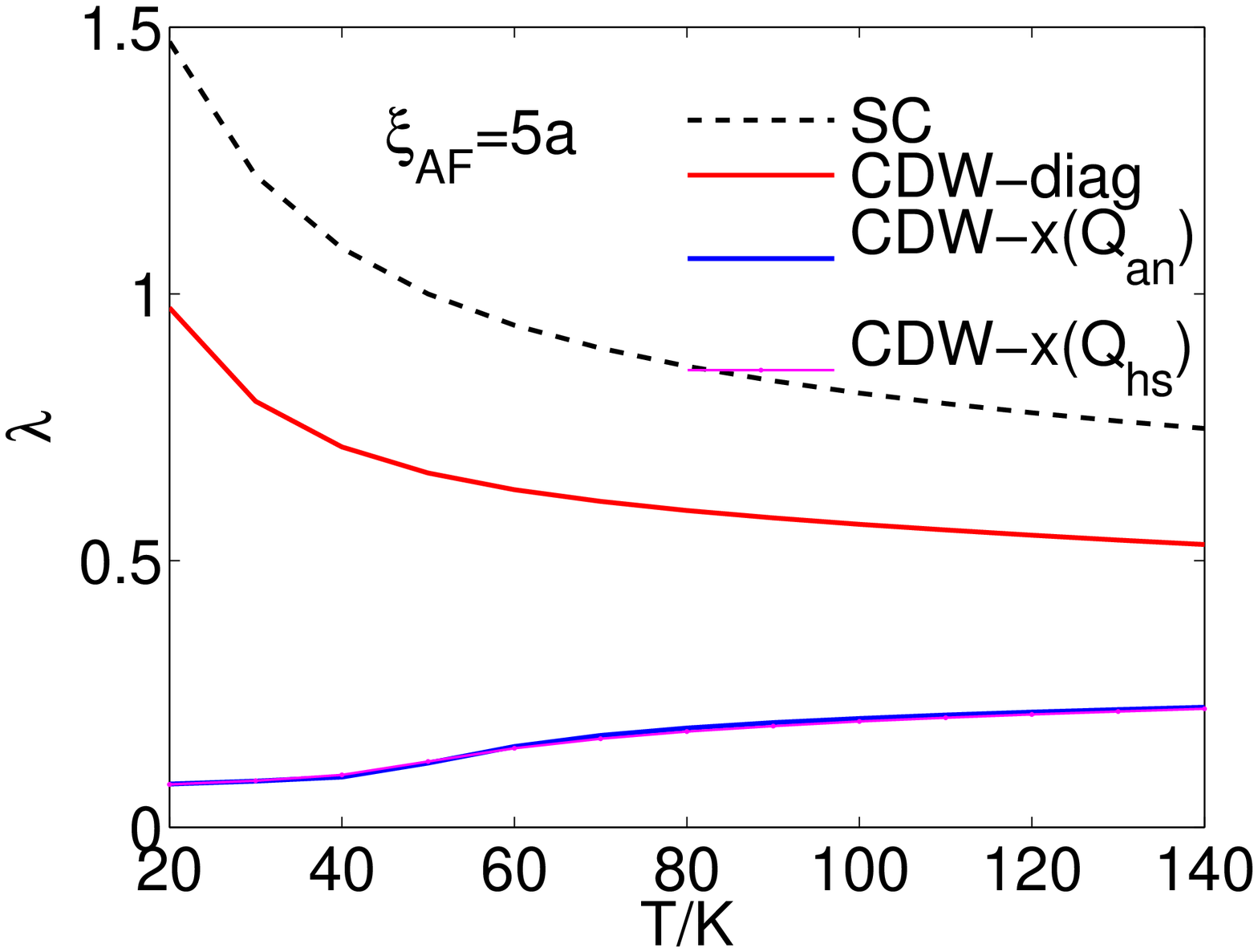}
\includegraphics[width=.95\columnwidth]{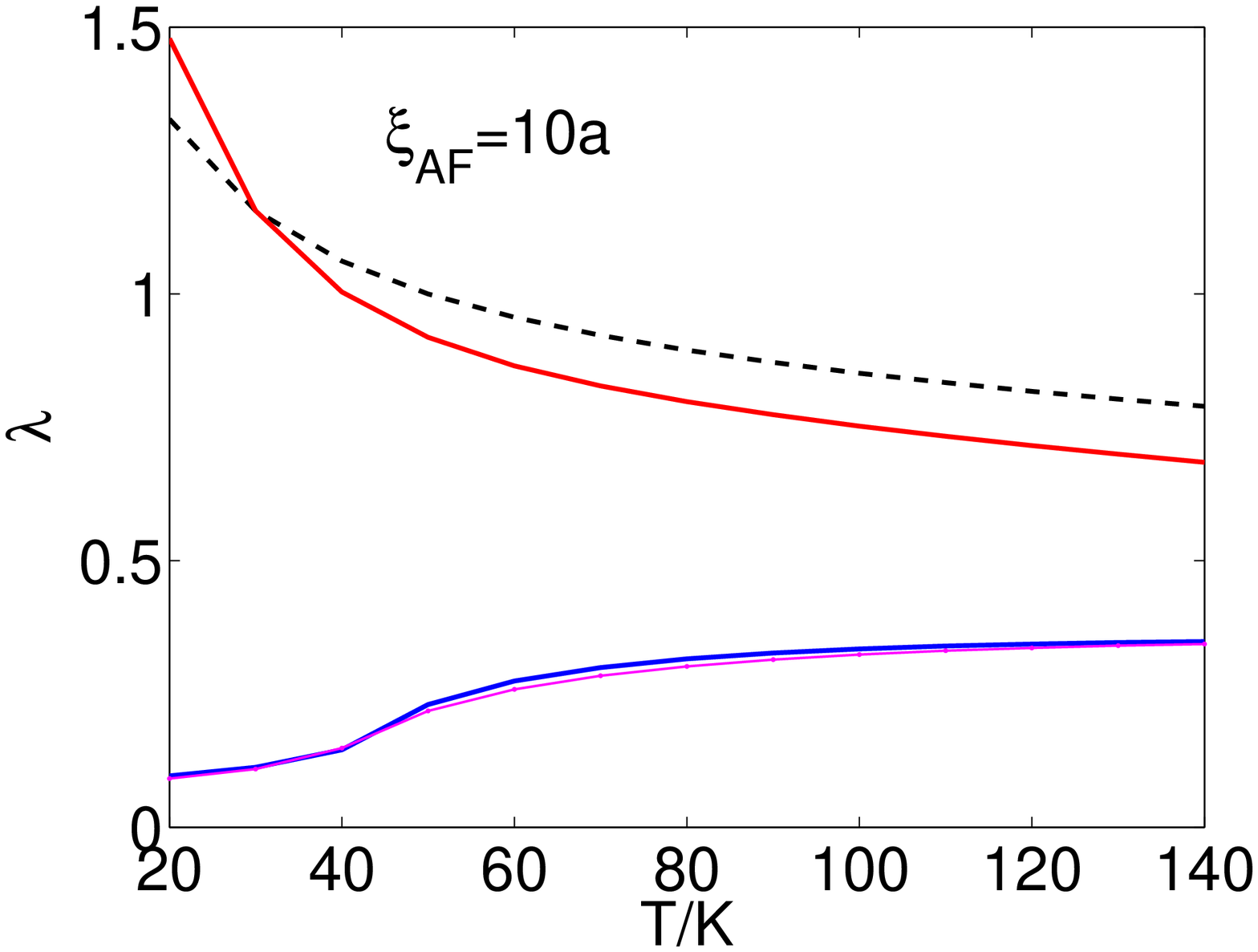}
\caption{Temperature dependence of the leading eigenvalues
for the superconducting (SC), diagonal 
CDW (CDW-diag) and bond-oriented CDW (CDW-x) states, for $\xi_{AF}=5a$ and $10a$,
in an approximation where $G$ is based on a renormalized dispersion taken from ARPES data.}
\label{Tdep:eigenvalue_sup}
\end{figure}
Supplementary Fig.~\ref{Tdep:eigenvalue_sup} shows the effect of larger antiferromagnetic
correlation lengths.
We use the same parameters for interactions as for $\xi_{AF}=2a$, except for the overall
prefactor $g^{2}_{sf}\chi_Q$, which we set to
0.49 eV and 0.35 eV for $\xi_{AF}=5a$ and $10a$, respectively, in order to obtain
a superconducting T$_c$ of 50 K.
As can be seen, increasing the antiferromagnetic
correlation length leads to similar results to those presented in the main text for $\xi_{AF}=2a$ (Fig.~2).

\section*{Self-energy effects}
\begin{figure}[tbph]
\includegraphics[width=.49\columnwidth]{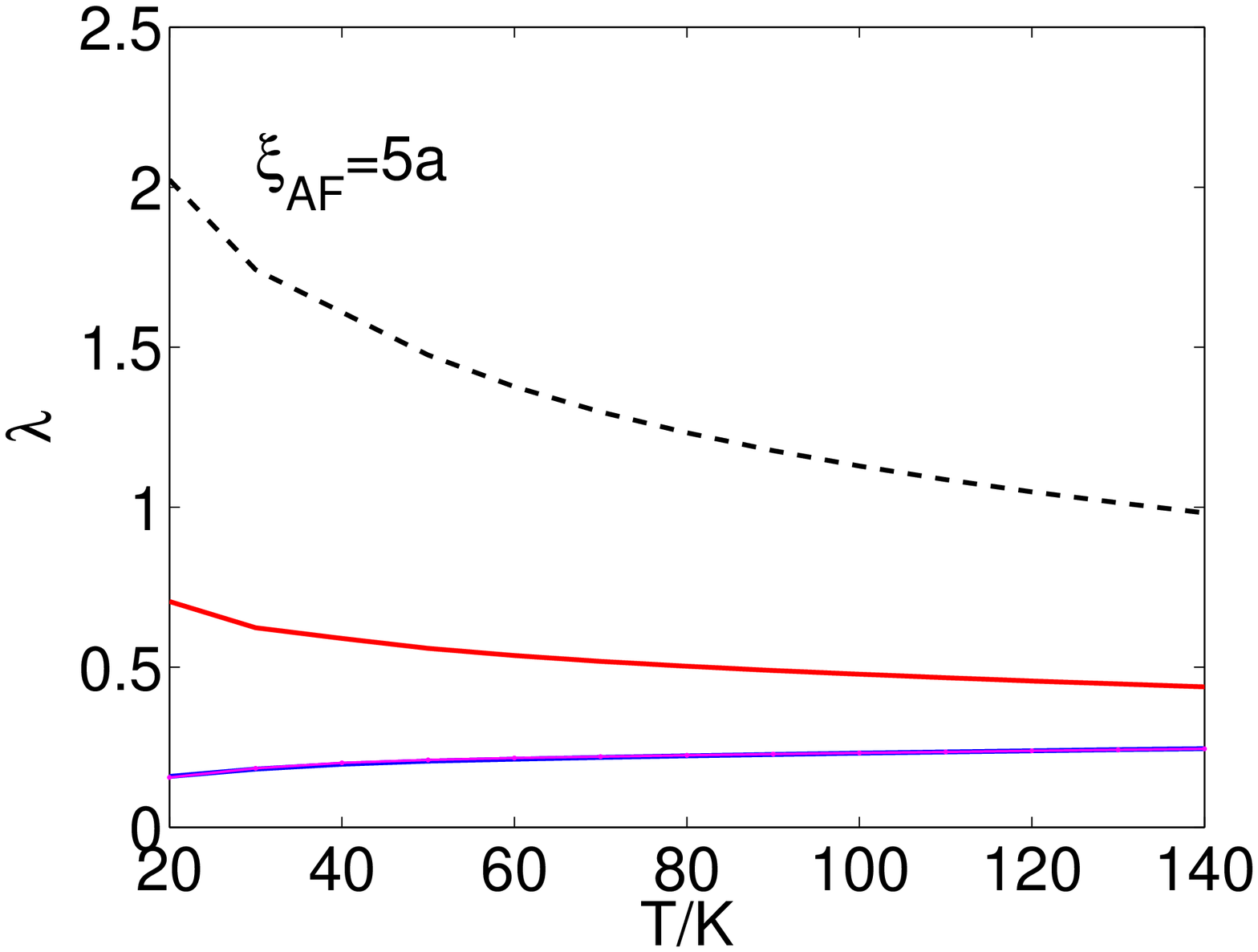}
\includegraphics[width=.49\columnwidth]{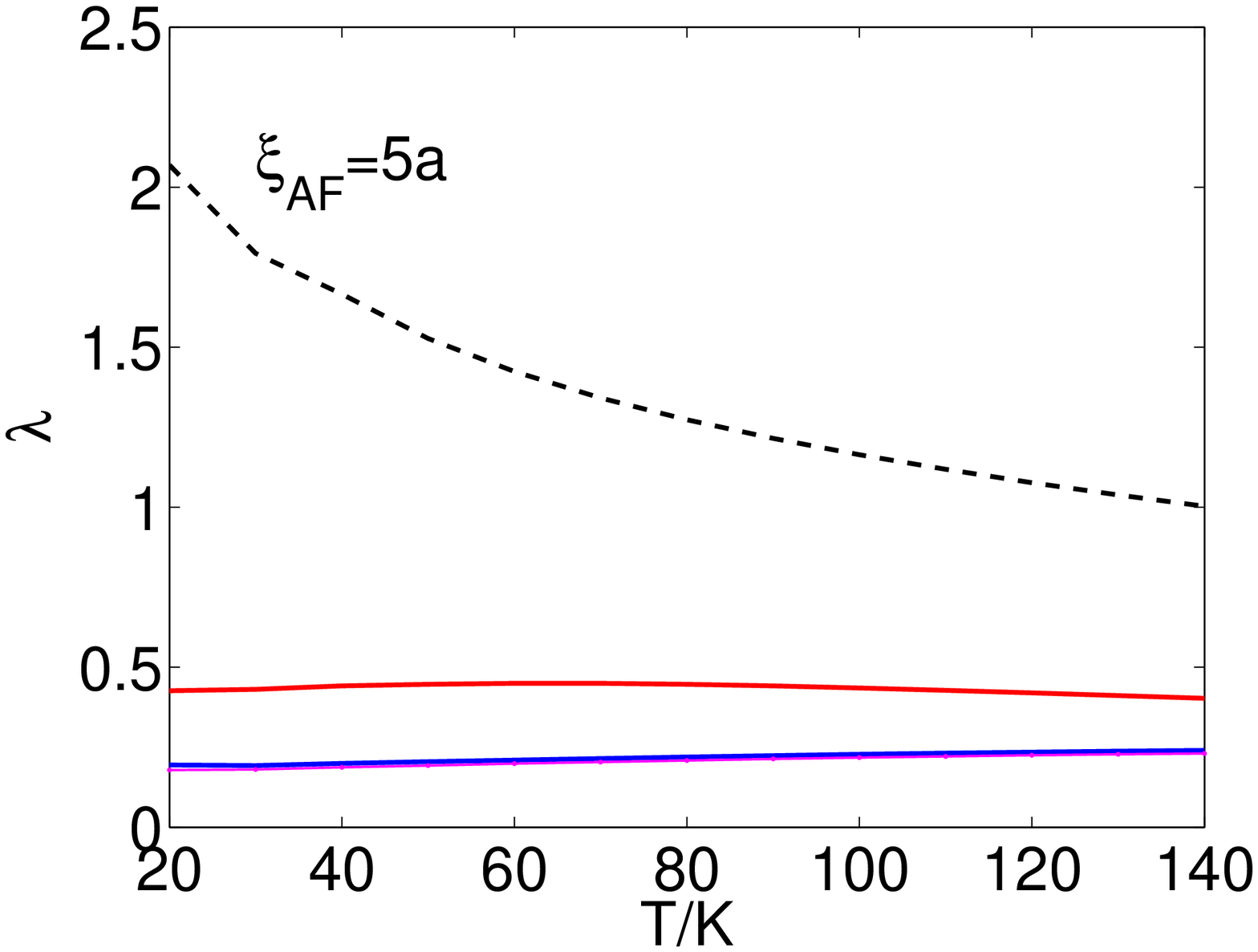}
\includegraphics[width=.49\columnwidth]{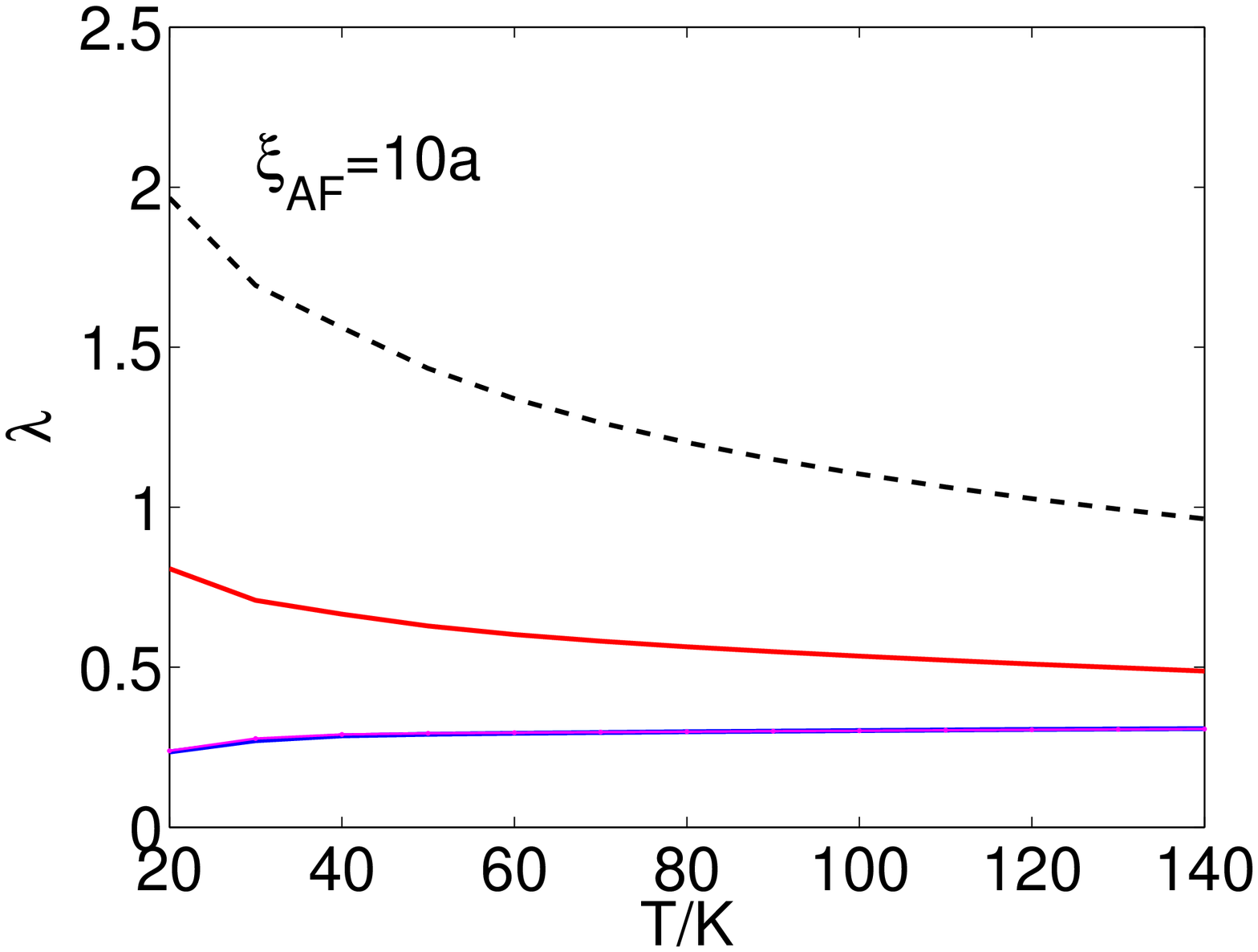}
\includegraphics[width=.49\columnwidth]{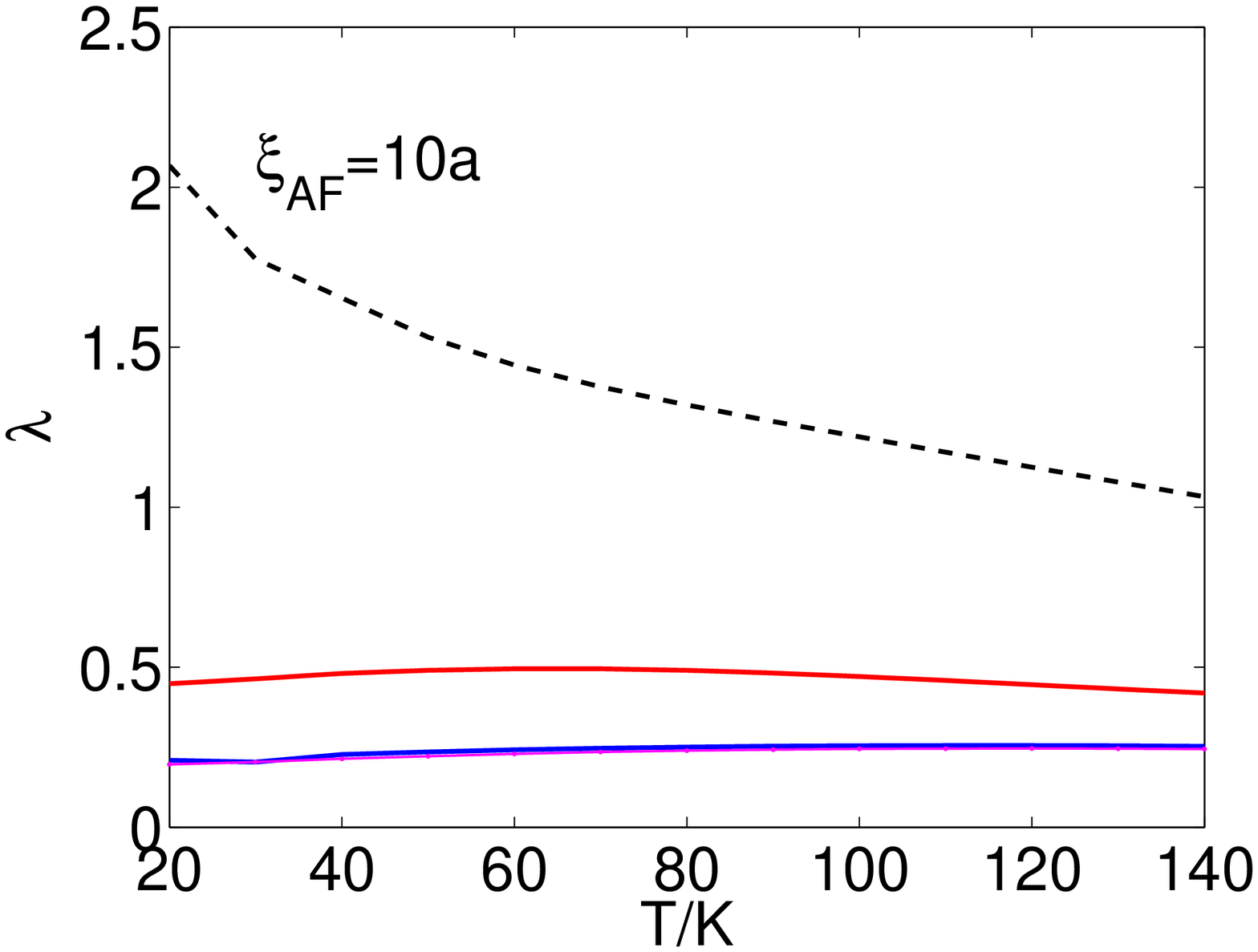}
\caption{Temperature dependence of the leading eigenvalues
for the superconducting (SC), diagonal 
CDW (CDW-diag) and bond-oriented CDW (CDW-x) states, for $\xi_{AF}=5a$ and $10a$,
obtained using dressed Green's functions. The left column shows 
the temperature dependence of the eigenvalues when the Fermi surface is kept unchanged in the full 
Green's function ($\xi_k$ dropped in the numerator of Eq.~6),
whereas for the figures in the right column, the full expression for $G$ in Eq.~6 is used.}
\label{Tdep:eigenvalue_sup2}
\end{figure}

Now, we include the effect of the fermion self-energy on the temperature dependence 
of the eigenvalues. Supplementary Fig.~\ref{Tdep:eigenvalue_sup2} shows the temperature
dependence of the eigenvalues for larger antiferromagnetic correlation lengths.
For this study we use $g^{2}_{sf}\chi_Q=3.2$eV and keep all other parameters the same as in the main text.
\section*{Effect of the band structure on the CDW order}\label{sec:comin}
\begin{figure}[h]
\includegraphics[width=0.95\columnwidth]{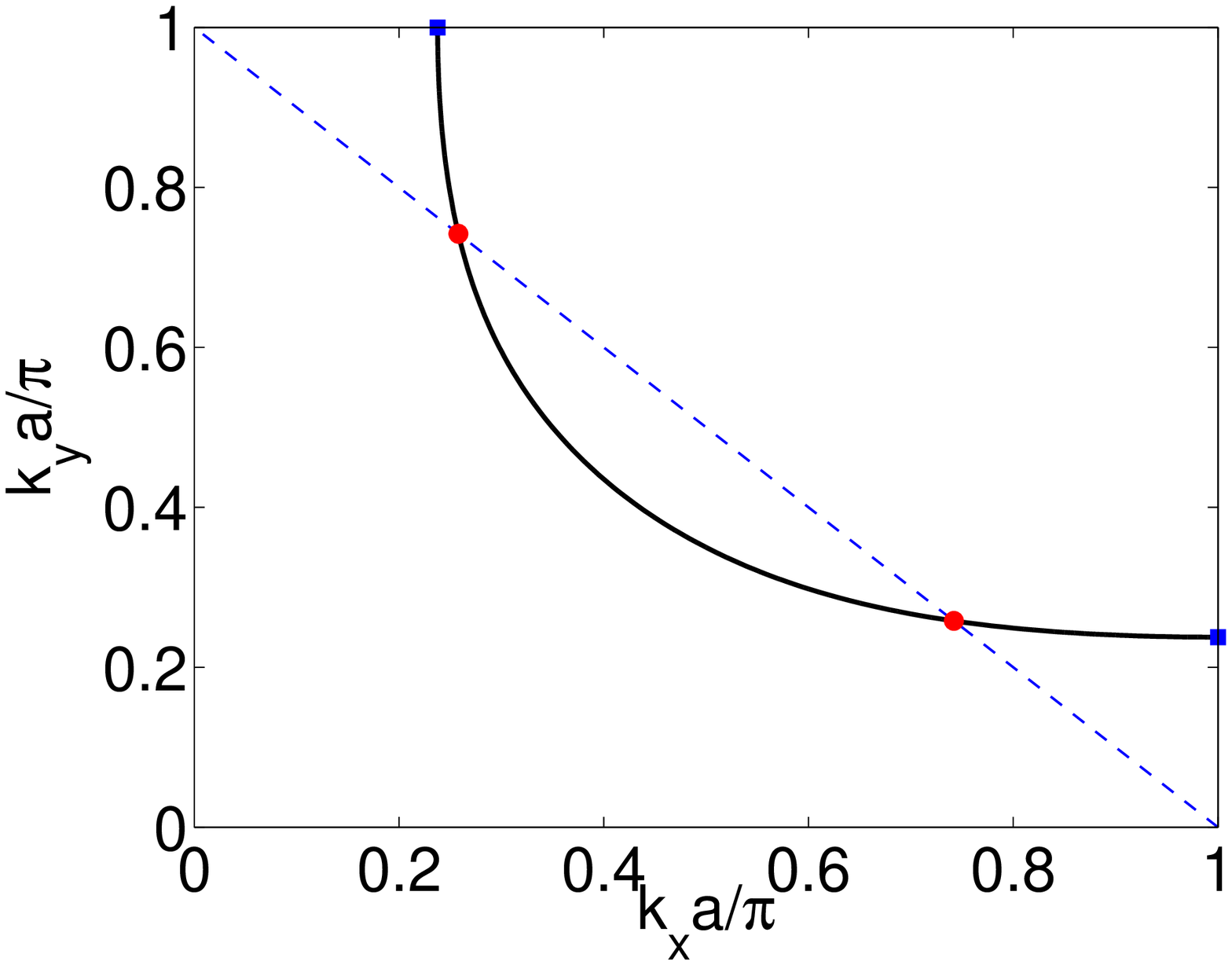}
\includegraphics[width=0.95\columnwidth]{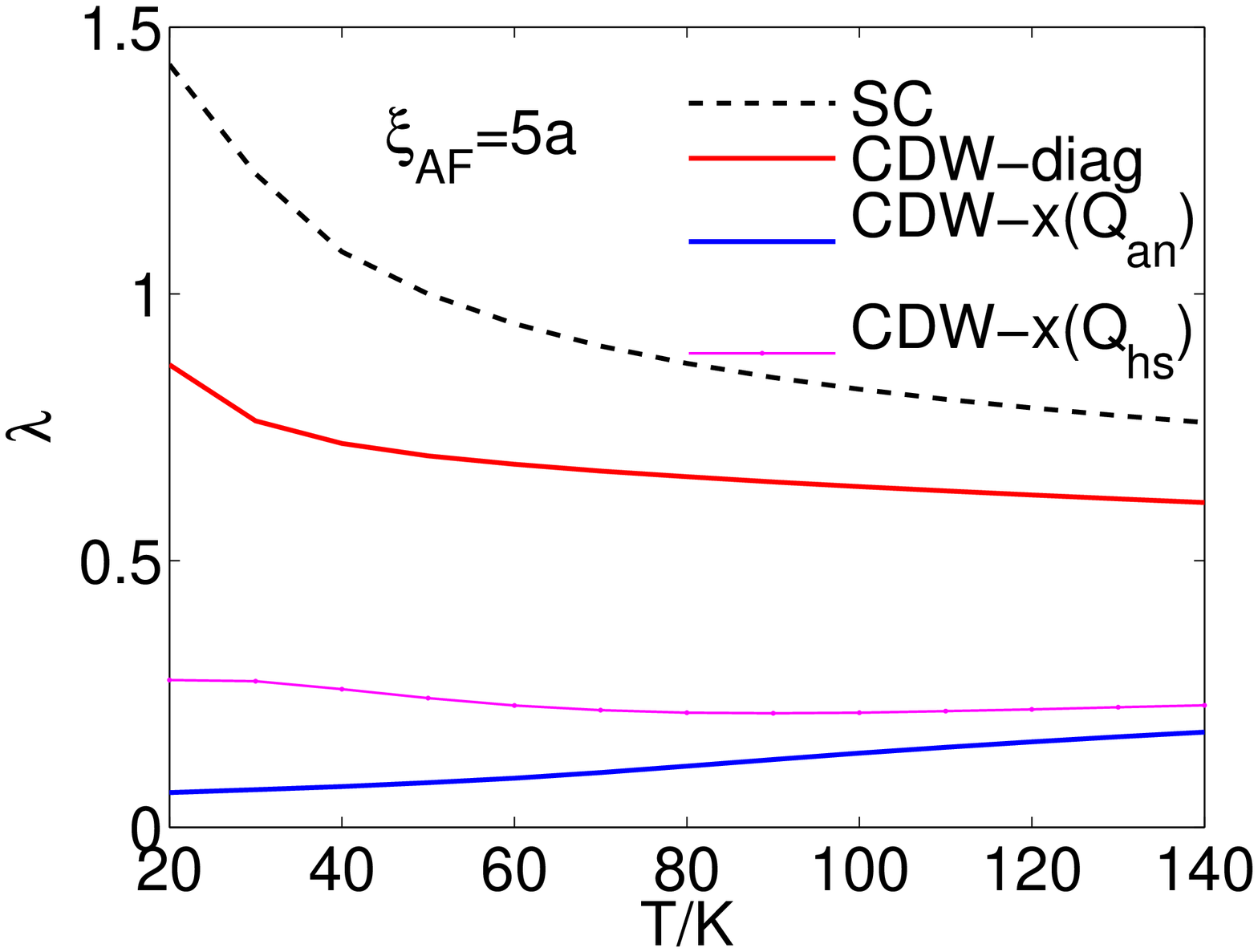}
\caption{Fermi surface for the alternate fermionic dispersion
is shown in the upper panel. The lower one shows the temperature dependence
of the leading eigenvalue for the superconducting (SC), diagonal 
CDW (CDW-diag), and bond-oriented CDW (CDW-x) states for ordering vectors $Q_{hs}$
and $Q_{an}$, in an approximation where $G$ is based on a renormalized dispersion taken from ARPES data.}
\label{fig:comin}
\end{figure}

We study a different fermionic dispersion where $Q_{an}$ and $Q_{hs}$ are further
separated than in the previous dispersion to test how this influences the solutions at $(Q_{an},0)$ and $(Q_{hs},0)$.
Supplementary Fig.~\ref{fig:comin} shows the Fermi surface for this alternate band structure
and the resulting temperature dependence of the leading eigenvalues
for various CDW states. Here we use $\xi_{AF}=5a$
and keep the rest of the parameters for the interaction as before,
except again we adjust $g^{2}_{sf}\chi_Q$ to 0.73 eV in order to obtain a superconducting T$_c$ of 50 K.
The fermionic dispersion is
\begin{eqnarray}
 \xi(k_x,k_y)&=& 0.4 \left( \cos k_xa + \cos k_ya \right) -0.32 \cos k_xa \cos k_ya \nonumber \\
 & & + 0.04  \left( \cos 2k_xa + \cos 2k_ya \right) - 0.15
 \label{eq:genericband}
\end{eqnarray}
where all energy scales are in eV. For this dispersion, we can clearly see that the
eigenvalues of the CDW-x state for $Q_{hs}$ and $Q_{an}$ are quite different, though neither
exhibit a log. We have also tested the tb1 and tb4 dispersions of Ref.~\onlinecite{MRN2007}
and do not find any qualitative difference in our conclusions.
\begin{figure}[tbph]
\includegraphics[width=0.95\columnwidth]{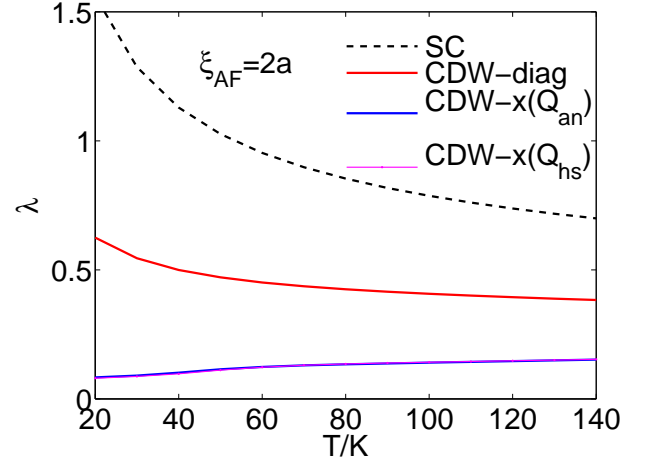}
\caption{The temperature dependence of the CDW eigenvalues with both spin-fluctuation and phonon mediated 
interactions, in an approximation where $G$ is based on a renormalized dispersion taken from ARPES data.
The dashed line is the eigenvalue for the superconducting state, which gets slightly enhanced with
the inclusion of the phonon mediated interaction.}
\label{Evec:phonon}
\end{figure}

\section*{Effect of the B$_{1g}$ phonon mode on the CDW order}\label{sec:phonon}
We consider the $B_{1g}$ phonon mode which involves an out of phase vibration
of adjacent oxygen atoms along the c-axis. In general, the electron-phonon matrix element is sensitive
to the details of the underlying electronic structure. We use the same fermion dispersion
as in the main text, and for the electron-phonon
matrix element we consider that based on a three-orbital model for the cuprates \cite{Johnston2010}.
The effective interaction reads (with $V_{sf}$ the spin-fluctuation mediated interaction of the main text)
\begin{eqnarray}
 V_{eff}&=& V_{sf} - g(k^\prime-Q/2,k-Q/2)g^*(k^\prime+Q/2,k+Q/2) \nonumber \\
 &\times& \frac{2\Omega_{B1g}}{\Omega^2+\Omega^2_{B1g}},
\end{eqnarray}
where $\Omega_{B1g}$ is the B$_{1g}$ phonon mode energy which we take to be 341 cm$^{-1}$, and
$g(k,k^\prime)$ is the electron-phonon matrix element which for this mode has a strong dependence
on $k$ and $k^\prime$ \cite{Johnston2010}.
For the overall magnitude of $g$ considered here, the Fermi velocity of 3.2 eV-{\AA}\cite{AVCMRN2004} along the zone diagonal from
our tight binding dispersion would be renormalized to 2.7 eV-{\AA}.
Results are show in Supplementary Fig.~4.
\section*{Convergence of eigenvalues}
\begin{figure}[h]
\includegraphics[width=0.95\columnwidth]{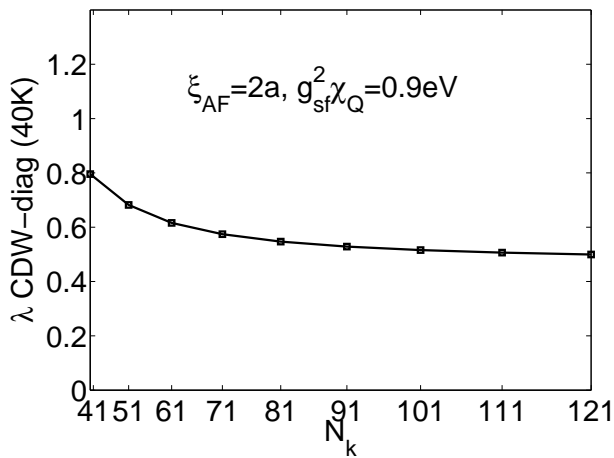}
\caption{The dependence of the maximum eigenvalue (as in Fig.~2) on the grid
size $N_k \times N_k$. These eigenvalues correspond to the CDW-diag state at 40 K.
In each case, we use 16 Matsubara frequencies.}
\label{fig:grid_check}
\end{figure}

We have performed our calculations on a 101$\times$101 momentum grid,
which corresponds to a 0.02$\pi/a$ momentum resolution. We checked
different grid sizes and find that this grid is sufficient for 
convergence of the eigenvalues. Supplementary Fig.~\ref{fig:grid_check} shows
the variation of eigenvalues as a function of the grid size.
We also find that the number of Matsubara frequencies used in
our calculation is sufficient for convergence of the eigenvalues
in the temperature range that we study.
\begin{figure}[h]
\includegraphics[width=0.95\columnwidth]{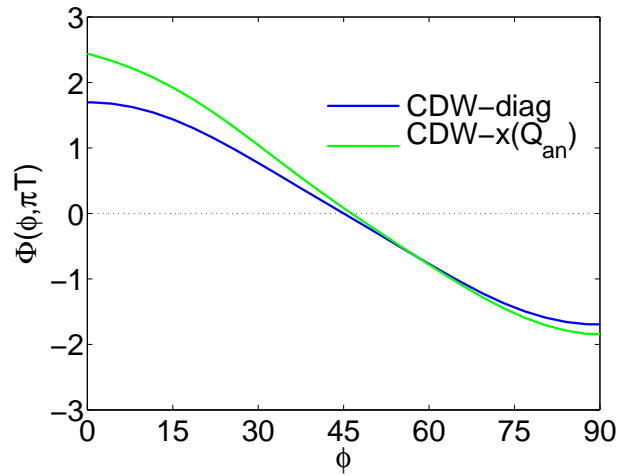}
\caption{The dependence of the eigenvector corresponding to the leading
eigenvalue (Fig.~2) for the ordering vectors ($Q_{\text{hs}},Q_{\text{hs}}$), and ($Q_{\text{an}},0$), 
projected on the Fermi surface, where $\phi=$0 corresponds to
the antinodal point. This is shown for the lowest Matsubara 
frequency at T$=$40 K.}
\label{Evec:FS}
\end{figure}
\section*{Eigenvector projection on the Fermi surface}\label{sec:evec_fsr}
In Supplementary Fig.~6 we show the projection of the eigenvector on the Fermi surface for the 
CDW-diag and CDW-x states. The full Brillouin zone dependence of these eigenvectors
are shown in the main text (Fig.~3). On the Fermi surface, the CDW-diag
state is well described by a $\cos 2\phi$ function, where $\phi$ is the angle along 
the Fermi surface. The bond oriented CDW-x state can be fit with $a+b\cos 2\phi + c \cos 4\phi$,
where these terms represent the $s$, $d$, and $s^\prime$ components, respectively, with
the $d$ component the dominant contribution.


\begin{thebibliography}{99}
\bibitem{foot1}
For lightly doped La based cuprates, diagonal charge order is seen.
\end{thebibliography}
\end{document}